\documentclass{ifacconf}

\usepackage{graphicx}      
\usepackage{natbib}        
\usepackage{amsmath}
\usepackage{tikz}
\usepackage{pgfplots}
\usepackage{stackengine}

\begin{document}
\begin{frontmatter}

\title{Optimal Freewheeling Control\\ of a Heavy-Duty Vehicle	Using\\ Mixed Integer Quadratic Programming} 


\author[First,Second]{Manne Held} 
\author[First]{Oscar Fl\"ardh}
\author[First]{Fredrik Roos} 
\author[Second]{Jonas M{\aa}rtensson}

\address[First]{Scania CV AB, 151\,87 S\"odert\"alje, Sweden}
\address[Second]{Division of Decision and Control Systems and the Integrated Transport Research Lab, KTH Royal Institute of Technology, 100\,44 Stockholm, Sweden.\\ e-mail: manneh@kth.se, oscar.flardh@scania.com, fredrik.roos@scania.com, jonas1@kth.se}

\begin{abstract}                
Improving the powertrain control of heavy-duty vehicles can be an efficient way to reduce the fuel consumption and thereby reduce both the operating cost and the environmental impact. One way of doing so is by using information about the upcoming driving conditions, known as look-ahead information, in order to coast with a gear engaged or to use freewheeling. Controllers using such techniques today mainly exist for vehicles in highway driving. This paper therefore targets how such control can be applied to vehicles with more variations in their velocity, such as distribution vehicles. The driving mission of such a vehicle is here formulated as an optimal control problem. The control variables are the tractive force, the braking force, and a Boolean variable representing closed or open powertrain. The problem is solved by a Model Predictive Controller, which at each iteration solves a Mixed Integer Quadratic Program. The fuel consumption is compared for four different control policies: a benchmark following the reference of the driving cycle, look-ahead control without freewheeling, freewheeling with the engine idling, and freewheeling with the engine turned off. Simulations on a driving cycle typically used for testing distribution vehicles show the potential of saving 10\,\%, 16\,\%, and 20\,\% respectively for the control policies compared with the benchmark, in all cases without increasing the trip time. 
\end{abstract}


\end{frontmatter}
\newcommand\blfootnote[1]{%
	\begingroup
	\renewcommand\thefootnote{}\footnote{#1}%
	\addtocounter{footnote}{-1}%
	\endgroup
}
\section{INTRODUCTION}
The \blfootnote{Funding provided by Swedish Governmental Agency for Innovation Systems ({VINNOVA}) through the {FFI} program is gratefully acknowledged.}road freight sector accounts for nearly 6\,\% of the total CO$_2$ emissions in the EU (\cite{TNO2015}). Therefore, the EU has agreed that emissions for new heavy-duty vehicles\,(HDVs) should be decreased compared to the 2019 level by 15\,\% in 2025 and by 30\,\% in 2030 (\cite{EuropeanParliment2018}). One way to decrease the emissions is to improve components, for instance by designing more efficient engines. Another way, which is the focus of this paper, is to improve the control of the vehicles by more fuel-efficient software.\par  
Many HDVs are today equipped with speed controllers using look-ahead information, such as road grade, to drive in a more fuel-efficient way. One example is Scania active prediction (\cite{Scania2012}), released in 2011, which can save 3\,\% fuel by adapting the speed profile to changes in the altitude. This is mainly done by coasting ahead of downhills in order to avoid braking. A few years later, freewheeling, i.e., decoupling the engine from the rest of the powertrain, was added to the controller. With this functionality, the fuel consumption was further reduced by 2\,\% (\cite{Scania2013}). Even though controllers such as these already exist for commercial use, their applications are limited to situations where the velocity only deviates by a few percent from a fixed set-point. This is typically the case for vehicles in highway driving. For distribution vehicles, for which the velocity has large variations however, such controllers do not exist to the same extent, which is one motivation for the work in this paper. \par 
\begin{figure}[]
	\centering
	\includegraphics[width=.99\columnwidth]{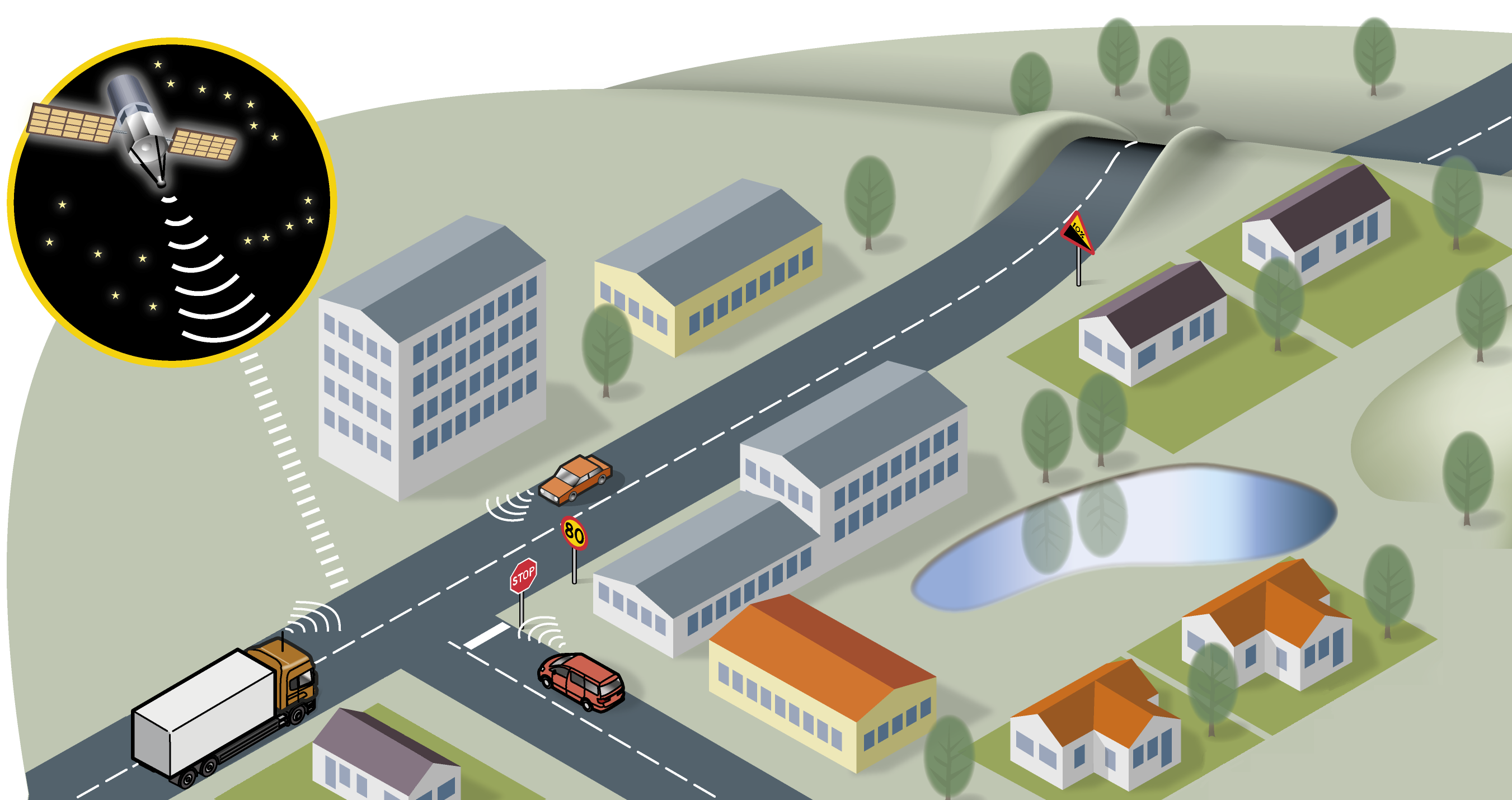}
	\caption{A heavy-duty vehicle in a driving mission involving varying velocity constraints and significant road slope.}
	\label{fig:drawing}
\end{figure}
The focus of this paper is fuel-efficient powertrain control of heavy-duty distribution vehicles. An example of such a scenario can be seen in Fig.~\ref{fig:drawing}. The vehicle has access to information about the curvature and the road grade of the upcoming downhill by using a map and GPS communication. It also has access to information about other traffic conditions such as the maximum speed restriction, either by using a map or by onboard sensors. While approaching both new speed restrictions and significant road grade, the vehicle can be controlled in different ways such as braking, coasting and freewheeling. For fuel efficient driving, braking should in most cases be avoided, but deciding when to use coasting and when to use freewheeling is not trivial. In addition, the decision may depend on whether the engine is idling or turned off during freewheeling. \par 
The approach for fuel reduction in this paper is, given a driving mission of a heavy-duty distribution vehicle, to formulate it as an Optimal Control Problem\,(OCP). This has been previously done in \cite{Held2018ITS}, where the problem was solved using both Pontryagin's Maximum Principle and Quadratic Programming\,(QP). The analysis was done from an energy perspective, and did not take the powertrain into account. Therefore, in this paper, a variable for whether the powertrain is closed or not is added to the QP-formulation. The main benefit of this is that the vehicle can save fuel by freewheeling with low engine speeds and thus reduce the drag losses in the engine. With the new variable, the problem becomes a Mixed Integer Quadratic Program\,(MIQP). \par 
Mixed integer programming has been used to solve similar problems before. A Model Predictive Controller\,(MPC) solving a mixed integer nonlinear program was applied to a heavy-duty vehicle in \cite{Kirches2011} in order to find the optimal gear shifting policy. The application was highway driving and freewheeling was not considered. Mixed integer programming has also been applied to the lateral movement of vehicles in order to avoid obstacles (\cite{Qian2016}). Another application was made for trains in \cite{Wang2011a}, where a mixed integer formulation was used in order to approximate nonlinear functions by a Piecewise Affine Function.\par 
One method for reducing the fuel consumption in HDVs is to reduce the total energy lost due to drag losses in the engine by decreasing the engine speed. This can be done by alternating between tractive power and freewheeling, known as Pulse-and-Glide\,(PnG). Four different cases of PnG were specifically studied in \cite{Xu2015}. One optimal cycle of PnG was performed for each case and they were then compared in terms of fuel consumption. External effects from road grade and varying velocity constraints, which might influence the timing of the PnG phases, are not considered. Another example is \cite{Li2011}, where PnG strategies are compared for different velocities in a car-following scenario. The engine drag torque at idling was not considered, and a continuous function could therefore be fitted to the fuel rate. \par 
One motivation for the work in this paper is the potential of decreasing the fuel consumption by reducing the drag losses in the engine, i.e., the losses caused by friction and pressure fall etc. For these kind of losses, the potential savings increase with decreasing gear numbers. This is because lower gear numbers mean more revolutions of the engine for the same driven distance. Distribution vehicles, as considered in this paper, drive at lower velocities and thus lower gear numbers than vehicles in highway applications. Therefore, reducing the drag losses is even more important for distribution vehicles compared to vehicles in highway driving, for which controllers using PnG already exist commercially. \par 
A similar problem to the one in this paper was solved in \cite{Henriksson2017b} using Dynamic Programming\,(DP). The main drawback of using DP is that the computation time can be very large due to the curse of dimensionality, i.e., the fact that the computation time grows exponentially with the number of states and control signals. Furthermore, the velocity in \cite{Henriksson2017b} is discretized and is thus no longer a continuous variable. \par 

The main contributions of this paper are:
\begin{enumerate}
	\item To find the optimal control of an HDV with the possibility to coast, freewheel with idle engine and freewheel with engine off, applied to a driving cycle with both significant road slope and varying velocity requirements.
	\item Compared to \cite{Henriksson2017b}, to solve the problem with continuous velocity, to avoid the curse of dimensionality and to investigate the effects of freewheeling with the engine off.  
\end{enumerate}
The outline of this paper is the following: The model of the vehicle and its engine is introduced in section \ref{sec:Modelling}. The OCP is presented in section \ref{sec:Problem_formulation} followed by simulation results in section \ref{sec:Simulation_results}, and conclusions in section \ref{sec:conclusion}.

\section{Modelling}\label{sec:Modelling}
In this section, the vehicle model is first described in terms of external forces, constraints, and time consumption. In section \ref{sec:DragLosses}, the drag losses in the engine are being modeled.

\subsection{Vehicle model}\label{sec:Vehicle_model_only}
Kinetic energy $K(s)=mv^2(s)/2$ is used as state variable as a function of position $s$, with $m$ being the vehicle mass and $v$ the velocity. The dynamics of the vehicle are given by
\begin{equation}\label{eq:d_K_d_s}
	\frac{dK(s)}{ds}=F_{fw}(s)+F_b(s)+F_a(K(s))+F_r(s)+F_g(s)
\end{equation}
where $F_{fw}(s)$ is the force at the flywheel, $F_b(s)$ the force caused by the brakes, $F_a(K(s))$ by the air resistance, $F_r(s)$ by the rolling resistance, and $F_g(s)$ by gravity. The resulting force on the flywheel is with closed powertrain given by 

\begin{equation}\label{eq:F_fw_closed}
	F_{fw}(s)= \begin{cases}
		F_{t}(s)-F_{d_{c}} & \text{powertrain closed}\\
		0 & \text{powertrain open}
	\end{cases}  
\end{equation} 
where $F_{t}(s)$ is the force generated by the combustion and $F_{d_{c}}$ is the drag force for closed powertrain which is discussed further in section \ref{sec:DragLosses}.

\begin{table}
	\caption{Parameters related to the vehicle and the environment.}
	\label{table:parameters}
	\begin{center}
		\begin{tabular}{|l|c|}
			\hline
			Parameter & Value\\ \hline
			$m$ - vehicle mass & 26\,000\,kg\\ 
			$r_w$ - wheel radius & 0.5\,m\\
			$c_d$ - air drag coefficient& 0.5 \\
			$\rho$ - air density & 1.292\,kg$\cdot$ m$^{-3}$ \\ 
			$A_f$ - vehicle cross-sectional area & 10\,m$^2$ \\
			$c_{r}$ - rolling resistance coefficient & 0.006 \\
			$\omega_{c}$ - engine speed powertrain closed & 1100\,RPM\\
			$\omega_{o}$ - engine speed powertrain open & \stackanchor{idle 500\,RPM}{engine off 0\,RPM} \\
			$J_{e}$ - moment of inertia engine & 4\,kg$\cdot$m$^2$\\
			$T_{d,0}$ - constant drag torque & \\ 
			$T_{d,1}$ - linear drag torque & \\ 
			\hline
		\end{tabular}
	\end{center}
\end{table}

In (\ref{eq:d_K_d_s}), $F_a(K(s))$ represents the air resistance such that
\begin{equation} \label{eq:F_a}
	F_a(K(s)) =-\frac{\rho A_{f} c_d K(s)}{m}
\end{equation}
where $\rho$ is the air density, $A_{f}$ is the vehicle frontal area, and $c_d$ is the air drag coefficient. The contribution from rolling resistance is given by
\begin{equation} \label{eq:F_r}
	F_r(s)=-mgc_{r}\cos(\alpha(s))	
\end{equation}
where $c_{r}$ is the coefficient for the rolling resistance, $g$ is the gravitational constant and $\alpha$ is the road slope. The gravitational force $F_g(s)$ is given by
\begin{equation} \label{eq:F_g}
	F_g(s)=-mg \sin(\alpha(s)).
\end{equation}
The brake force in (\ref{eq:d_K_d_s}) is constrained by
\begin{equation}
	-F_{b_{\text{max}}}\leq  F_{b} \leq 0.
\end{equation}
By writing the inverse velocity as
\begin{equation}\label{eq:velocity_inverse}
\frac{1}{v}=\sqrt{\frac{m}{2}}K^{-1/2},
\end{equation}
the tractive force is in addition to (\ref{eq:F_t_max}) constrained by the maximum tractive power $P_{\text{max}}$ as
\begin{equation}\label{eq:P_max}
	F_{fw} \leq P_{\text{max}}\sqrt{\frac{m}{2}}K^{-1/2}.
\end{equation}
By driving a distance $\text{d}s$ with velocity $v$, the consumed time $\text{d}t$ using (\ref{eq:velocity_inverse}) becomes
\begin{equation}\label{eq:dt_equals_ds_v}
	\text{d}t=\text{d}s\sqrt{\frac{m}{2}}K^{-1/2}.
\end{equation}

\subsection{Engine drag losses}\label{sec:DragLosses}
%
The energy losses due to engine drag are modelled as a force $F_{d}$ in (\ref{eq:F_fw_closed}). It can be calculated using the relation
\begin{equation}\label{eq:E_s_equals_P_v}
F_{d}=\frac{P}{v}
\end{equation}
where the power $P$ is calculated from the engine drag torque $T_{d}$ and engine speed $\omega$ as
\begin{equation}
P=T_d(\omega) \omega.
\end{equation}
The drag torque can be modelled to be linear in engine speed such that  	
\begin{equation} \label{eq:T_drag}
T_d(\omega)= T_{d,0}+T_{d,1}\omega
\end{equation} 
where $T_{d,0}$ and $T_{d,1}$ are found using least squares fit to experimental values from a Scania engine. Combining (\ref{eq:E_s_equals_P_v})-(\ref{eq:T_drag}) gives
\begin{equation} \label{eq:E_s}
F_{d}= \begin{cases}
\left(T_{d,0}+T_{d,1}\omega_{c} \right)\omega_{c}/v & \text{powertrain closed}\\
\left(T_{d,0}+T_{d,1}\omega_{o}\right)\omega_{o}/v & \text{powertrain open}\\
0 & \text{engine off}\\
\end{cases} 
\end{equation}
where $\omega_{c}$ is the engine speed with closed powertrain and $\omega_{o}$ is the engine speed with open powertrain. These are both set to constant values. For $\omega_{c}$, this is a simplification since it varies continuously between gear changes. The range of typically used engine speeds for an HDV is about 800-1500\,RPM and thus much smaller than the range used by engines in personal cars. The chosen value of $\omega_{c}=$1100\,RPM is a commonly used and efficient engine speed.

The Boolean variable $z\in \{0,1\}$ is introduced such that it attains the value $1$ if the powertrain is closed and $0$ if the powertrain is open. The tractive force is non-negative and can attain values up to its maximum $F_{t_{\text{max}}}$ only if the powertrain is closed. If the constraint
\begin{equation}\label{eq:F_t_max}
0 \leq F_{t}(s) \leq F_{t_{\text{max}}}z
\end{equation} 
is introduced, then (\ref{eq:F_fw_closed}) can be written
\begin{equation}\label{eq:F_fw}
F_{fw}(s) =  F_{t}(s)-F_{d_{c}}z.
\end{equation}

If the powertrain is open, the engine is either turned off or runs at idle. The energy needed for this is represented by the constant drag force $F_{d_{o}}$, which is taken into account in the cost function of the OCP. \par 
The drag forces $F_{d_{c}}$ and $F_{d_{o}}$ are calculated using the relation
\begin{equation}
F_d=\omega T_d(\omega)\sqrt{\frac{m}{2}}K^{-1/2}
\end{equation}
where the engine drag torque is modelled as linear in engine speed as in (\ref{eq:T_drag}).\par 

It should be noted that freewheeling with idle engine will change the working points of the engine. However, the combustion efficiency, i.e., the efficiency considering the engine drag as part of the useful energy, of a diesel engine is not very different when idling compared to other working points. Therefore, the conversion ratio between fuel and energy is modelled to be independent of engine speed and torque.

\section{Problem formulation}\label{sec:Problem_formulation}
This section first describes how the constraints on velocity are set based on the driving cycle and statistics from real HDV operations. Next, the problem is formulated and solved as an MIQP. 

\subsection{Velocity constraints}
The driving cycle used for the simulations in this paper is based on a cycle used at Scania CV AB for testing distribution vehicles. The cycle contains the road grade and a piecewise constant reference speed. The original cycle contains a few stretches where the velocity is constant for more than 1\,km. Such stretches are here reduced to be only 1\,km long. The motivation for this removal is that look-ahead control with constant velocity profile is already a well-studied topics, see for instance \cite{Hellstrom2007a}.\par 
The constraints on the velocity are set based on the method introduced in \cite{Held2018ITS}. In that paper, a statistical analysis was performed to find average and standard deviation of decelerations of HDVs for different start and end velocities. Now in this paper, these data are fitted to a 2-dimensional polynomial using a least squares method such that the average deceleration $d_{\mu}(v_{1},v_{2})$ in m/s$^2$ when decelerating from $v_1$ to $v_{2}$ in m/s is given by
\begin{equation}
	\begin{split}
		d_{\mu}(v_1,v_2)=0.366 + 0.0771 v_1 -0.0849 v_{2} \\
		-0.00185 v_1^2 + 0.00348 v_1 v_{2} -0.00214 v_{2}^2	
	\end{split}
\end{equation}
and the corresponding standard deviation $\Sigma(v_1,v_2)$ is given by 
\begin{equation}
	\begin{split}
		\Sigma(v_{1},v_{2})= 0.187 + 0.0250 v_1 -0.0327 v_2 \\
		-0.000734 v_1^2 + 0.00187 v_1 v_2 -0.00101 v_2^2
	\end{split}
\end{equation}

\begin{figure}
	\centering
	\input{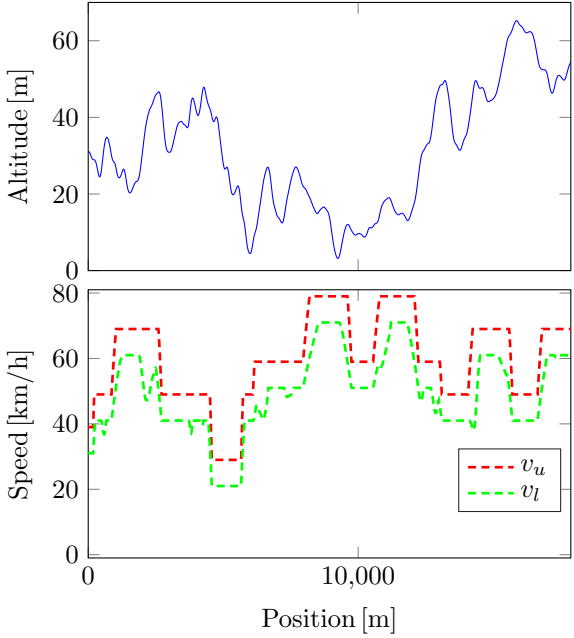}
	\caption{The altitude of the driving cycle in the top figure and the velocity corridor in the bottom figure.}
	\label{fig:driving_cycle}
\end{figure}
The parameters used in the vehicle model can be seen in Table~\ref{table:parameters}.
These functions are used to create a velocity corridor, i.e., a lower and an upper constraint $v_l$ and $v_u$ for the velocity. Following the methods developed in \cite{Held2018ITS}, the procedure can be summarized as follows:
\begin{enumerate}
	\item Starting with a piecewise constant velocity reference $v_{ref}$
	\item Set $v_l=v_{\text{ref}}-\Delta v$
	\item Set $v_u=v_{\text{ref}}+\Delta v$
	\item For each deceleration from $v_1$ to $v_2$ set $v_l$ according to the deceleration $d_{\mu}(v_1,v_2)-n_{\Sigma}\Sigma(v_{1},v_{2})$ and $v_u$ according to the deceleration $d_{\mu}(v_1,v_2)+n_{\Sigma}\Sigma(v_{1},v_{2})$
	\item For each acceleration, set $v_l$ and $v_u$ such that they follow the constant acceleration $a_l$ and $a_u$ respectively, given in Table \ref{table:driving_corridor}.
\end{enumerate}
Where $\Delta v$ and $n_{\Sigma}$ are settings for the width of the corridor in terms of deviation during constant velocity and during deceleration respectively.\par

The road grade of the driving cycle has a maximum inclination of 4.3\,\% in an uphill. Such sections can possibly reduce the velocity of the HDV significantly. Setting the lower velocity constraint directly as above might therefore result in infeasibility. To mitigate this, the lower constraint is modified in order to always contain a feasible solution. This is done by discretizing the constraint with some small step size and for each step $k$ setting 
\begin{equation}
	v_l[k]= \min \left(v_l[k], v_l[k-1]+\Delta v_l \right)
\end{equation}
where $\Delta v_l$ is the acceleration yielded by maximum tractive power. The altitude and the resulting velocity constraints can be seen in Fig.~\ref{fig:driving_cycle}.

\begin{table}
	\begin{center}
		\caption{Settings for creating the velocity corridor.}
		\label{table:driving_corridor}
		\begin{tabular}{ |l|l|l|}
			\hline
			& Benchmark & wider  \\ \hline
			$\Delta v$\,[km/h]& 1 & 4 \\ \hline
			$n_{\Sigma}$\,[-] & 0.5 & 1 \\ \hline
			$a_l$\,[m/s$^2$] & 0.3 & 0.25 \\ \hline
			$a_u$\,[m/s$^2$] & 0.4 & 0.6 \\ \hline
		\end{tabular}
	\end{center}
\end{table}

\subsection{Mixed Integer Quadratic Program}
The problem is discretized with $\Delta s = 15$\,m using zero order hold and formulated as an MIQP. In order to do this, the cost function needs to be quadratic in the continuous state and control variables, the constraints need to be linear, and a continuous variable cannot be multiplied by the Boolean variable. The time consumption (\ref{eq:dt_equals_ds_v}) is used in the cost function while (\ref{eq:P_max}) and (\ref{eq:F_fw}) are used as constraints. They all contain the expression $K^{-1/2}$, and need to be approximated by the second, first, and zeroth order Taylor approximation respectively.\par 
The second order Taylor approximation of the kinetic energy around a reference trajectory $K_r$ is given by
\begin{equation}\label{eq:K_Taylor}
	\begin{split}
		K^{-1/2}\approx K_r^{-1/2} &- \frac{1}{2}K_r^{-3/2}(K-K_r)\\		
		&+ \frac{3}{8}K_r^{-5/2}(K-K_r)^2.
	\end{split}
\end{equation}
The Taylor approximation of different degrees at step $k$ become
\begin{equation} \label{eq:K_inv}
	K_{k}^{-1/2}\approx \begin{cases}
		\theta_{0,k}+\theta_{1,k}K_k+\theta_{2,k}K_k^{2} & \text{second order}\\
		\phi_{0,k}+\phi_{1,k}K_k & \text{first order}\\
		\varphi_{0,k} & \text{zeroth order}\\
	\end{cases} 
\end{equation}
where the second order coefficients are given by
\begin{subequations}		
	\begin{align}
		\theta_{0,k}&=\ \ \frac{15}{8}\sqrt{\frac{m}{2}}K_{r,k}^{-1/2}\\
		\theta_{1,k}&=-\frac{10}{8}\sqrt{\frac{m}{2}}K_{r,k}^{-3/2}\\
		\theta_{2,k}&=\ \ \frac{3}{8}\sqrt{\frac{m}{2}}K_{r,k}^{-5/2},
	\end{align}
\end{subequations}
the first order coefficients are given by
\begin{subequations}
	\begin{align}
		\phi_{0,k}&=\ \ \frac{3}{2}\sqrt{\frac{m}{2}}K_{r,k}^{-1/2}\\
		\phi_{1,k}&=-\frac{1}{2}\sqrt{\frac{m}{2}}K_{r,k}^{-3/2},
	\end{align}
\end{subequations}
and the zeroth order coefficient is given by
\begin{equation}
	\varphi_{0,k}=\sqrt{\frac{m}{2}}K_{r,k}^{-1/2}.
\end{equation}\par 
The OCP is solved in a receding horizon approach using an MPC with a control horizon of $N_H=60$\,steps. This leads to a control horizon with distance $\Delta s N_H=900$\,m which is enough for reaching the optimum value within a few parts per thousand (\cite{Held2018ITS}). Solving the OCP using an MPC instead of offline as an optimization problem is motivated by the MIQP-solver not converging to a solution when solving over the full driving distance.
For each discretized step $k$, the problem is formulated as an MIQP as:
\begin{subequations}
	\label{eq:problem_formulated}
	\begin{align}
		&\underset{F_{t},z}{\text{min}}  & &\sum_{j=k}^{k+N_H-1} & &\hspace{-5.3cm}\Delta s\left(F_{t,j}+\omega_oT_d(\omega_o)\varphi_{0,j}(1-z_j)\right)\label{eq:costfuncion_tractive}\\
		& & & & &\hspace{-5.3cm}+\beta_{g} \Delta z_{j}\label{eq:costfuncion_gear}\\
		& & & & &\hspace{-5.3cm} +\beta_t\Delta s \left(\theta_{0,j}+\theta_{1,j}K_j+\theta_{2,j}K_j^{2}\right)\label{eq:costfuncion_time} \\
		&&&&&\hspace{-5.3cm}-K_{N_H}\label{eq:costfuncion_rest}\\
		&\text{s.t.} & & \hspace{-0.3cm}K_{j+1}=AK_j+B(F_{t,j}-F_{d_{c},j}z_j\, +F_{b,j})+w_j\label{eq:problem_formulated_dyn}\\
		&&& \hspace{-0.3cm}F_{d_c,j}=\omega_{c} T_d(\omega_c)\left( \phi_{0,j}+\phi_{1,j}K_j\right)\label{eq:problem_formulated_drag}\\
		&&& \hspace{-0.3cm}K_{l,j} \leq K_j\leq K_{u,j}\label{eq:problem_formulated_corridor}\\
		&&& \hspace{-0.3cm}F_{t,j}\leq P_{\text{max}}\left( \phi_{0,j}+  \phi_{1,j}K_j\right)\label{eq:problem_formulated_P}\\
		&&& \hspace{-0.3cm}0\leq F_{t,j} \leq z_j F_{t_{\text{max}}}\label{eq:problem_formulated_Ft}\\
		&&& \hspace{-0.3cm}-F_{b_{\text{max}}} \leq F_{b,j} \leq 0\label{eq:problem_formulated_Fb} \\
		&&& \hspace{-0.3cm}K_k \text{ given} 
	\end{align}
\end{subequations}
where $A, B$ and $w_j$ are given by

\begin{subequations}\label{eq:A_B_v}
	\begin{align}
		A&=e^{-A_c\Delta s},\\
		B&=\frac{1}{A_c}(1-A)\\
		w_j&= -Bmg\left(
		\sin\alpha_j+c_{r}\cos\alpha_j \right)
	\end{align}
\end{subequations}
where $A_c$ is the state dependent coefficient in the continuous model (\ref{eq:d_K_d_s}) which is given by (\ref{eq:F_a}) as

\begin{equation}
	A_{c} = -\frac{\rho A_{f} c_d}{m}. 	  
\end{equation}

The cost function is the sum of the energy used for traction and idling (\ref{eq:costfuncion_tractive}), the cost for gear changes (\ref{eq:costfuncion_gear}) with $\Delta z_{j}=|z_{j+1}-z_{j}|$, and the cost for time consumption (\ref{eq:costfuncion_time}). The kinetic energy at the end of the horizon (\ref{eq:costfuncion_rest}) is added such that it is not always optimal to coast at the end of the horizon. The constraints consist of the dynamics of the vehicle (\ref{eq:problem_formulated_dyn}) with the drag force given by (\ref{eq:problem_formulated_drag}), the constraints on velocity (\ref{eq:problem_formulated_corridor}), maximum tractive power (\ref{eq:problem_formulated_P}), maximum tractive force (\ref{eq:problem_formulated_Ft}), and maximum braking force (\ref{eq:problem_formulated_Fb}). The constant $\beta_{g}$ in (\ref{eq:costfuncion_gear}) is a penalty for engaging or disengaging a gear. The motivation for this penalty is that when again engaging a gear, the rotational speed of the engine must be increased such that an amount of energy corresponding to the difference in rotational energy is consumed. Since both engaging and disengaging a gear is penalized in (\ref{eq:costfuncion_gear}), $\beta_g$ is split in two such that
\begin{equation}
	\beta_g = \frac{1}{2}J_e \frac{\omega_c^2-\omega_o^2}{2}
\end{equation}
where $J_e$ is the engine moment of inertia. The penalty for time consumption $\beta_t$ in (\ref{eq:costfuncion_time}) is set such that the different control policies obtain similar trip times in order to make a fair comparison of their energy consumption.

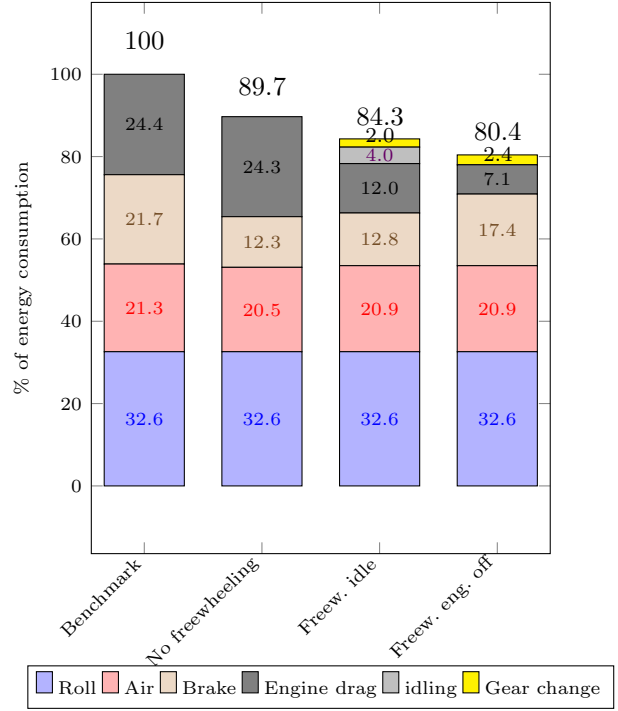
\begin{figure}
	\begin{tikzpicture}
\definecolor{x11gray}{rgb}{0.75, 0.75, 0.75}
\pgfplotsset{width=3in,height=3.5in,compat=1.9}
\begin{axis}[
ybar stacked,
bar width=30pt,
nodes near coords,
enlargelimits=0.15,
legend style={at={(0.5,-0.22)},
	anchor=north,legend columns=-1, yshift=0.05in},
ylabel={\% of energy consumption},
symbolic x coords={tool1, tool2, tool3, tool4},
style={font=\scriptsize},
xtick=data,
ymin=-1,
ymax=102,
xticklabels={Benchmark,No freewheeling,Freew. idle,Freew. eng. off},
x tick label style={rotate=45,anchor=east},
every node near coord/.append style = {
	/pgf/number format/.cd,
	fixed,
	fixed zerofill,
	precision=1,
},
]

\addplot+[ybar,draw=black] plot coordinates {(tool1,32.6) (tool2,32.6) 
	(tool3,32.6) (tool4,32.6) };
\addplot+[ybar,draw=black] plot coordinates {(tool1,21.3) (tool2,20.5)
	(tool3,20.9) (tool4,20.9)};
\addplot+[ybar,draw=black] plot coordinates {(tool1,21.7) (tool2,12.3) 
	(tool3,12.8) (tool4,17.4) };
\addplot+[ybar,draw=black] plot coordinates {(tool1,24.4) (tool2,24.3) 
	(tool3,12.0) (tool4,7.11)};
\addplot+[ybar,fill= x11gray,draw=black] plot coordinates {(tool1,0) (tool2,0) 
	(tool3,4.0) (tool4,0)};
\addplot+[ybar,fill= yellow,draw=black,nodes near coords = {%
	\pgfmathprintnumberto[fixed,assume math mode=true]{\pgfplotspointmeta}{\myval}%
	\pgfmathparse{\myval<10?:\myval}\pgfmathresult%
} ] plot coordinates {(tool1,0) (tool2,0) 
	(tool3,2.0) (tool4,2.4) };
\legend{ \strut Roll, \strut Air,\strut Brake,\strut Engine drag,\strut idling,\strut Gear change}
\end{axis}

\node[draw=none, font=\footnotesize] at (3.80,5.55){2.0 };
\node[draw=none, font=\footnotesize] at (5.35,5.30){2.4 };

\node[draw=none, font=\normalsize] at (0.70,6.8){100 };
\node[draw=none, font=\normalsize] at (2.25,6.2){89.7 };
\node[draw=none, font=\normalsize] at (3.80,5.8){84.3 };
\node[draw=none, font=\normalsize] at (5.35,5.6){80.4 };
\end{tikzpicture}
	\caption{Energy losses divided into categories for the different control policies.}
	\label{fig:source_DP}
\end{figure}

\section{Simulation results}\label{sec:Simulation_results}
Simulations were performed in Matlab using the toolbox Yalmip (\cite{Lofberg2004}) with the solver Gurobi (\cite{Optimization2018}).
Four different control policies were simulated in order to compare their fuel consumption:
\begin{enumerate}
	\item Benchmark, no freewheeling,
	\item No freewheeling,
	\item Freewheeling with idling engine,
	\item Freewheeling with engine off.
\end{enumerate}
For the first policy, the benchmark velocity corridor from Table~\ref{table:driving_corridor} was chosen. For the last three policies, the wide velocity corridor was chosen.\par 
The normalized fuel consumption can be seen in Fig.~\ref{fig:source_DP}. The fuel consumption is split into the parts originating from rolling resistance, air resistance, braking, engine drag, idling and gear changes. As can be seen, the losses due to rolling resistance is the same for all control policies and the losses due to air resistance only have small deviations. The savings by using a wider velocity corridor found by comparing policy 1 and 2 comes from reduction of the losses due to braking, as found in \cite{Held2018ITS}. \par 
\begin{table}
	\begin{center}
		\caption{Resulting energy consumption and trip time as percentage of the benchmark for the different control policies.}
		\label{table:result}
		\begin{tabular}{ |l|l|l|l|l|}
			\hline
			& Bench. & No freew. & Idling  & Eng. off  \\ \hline
			Energy\,[\%]& 100 & 89.7 & 84.3 & 80.4 \\ \hline
			Time\,[\%] 		& 100 & 99.4 & 99.7 & 99.7 \\ \hline
		\end{tabular}
	\end{center}
\end{table}

\begin{figure*}
	\centering
	\resizebox{\textwidth}{!}{\input{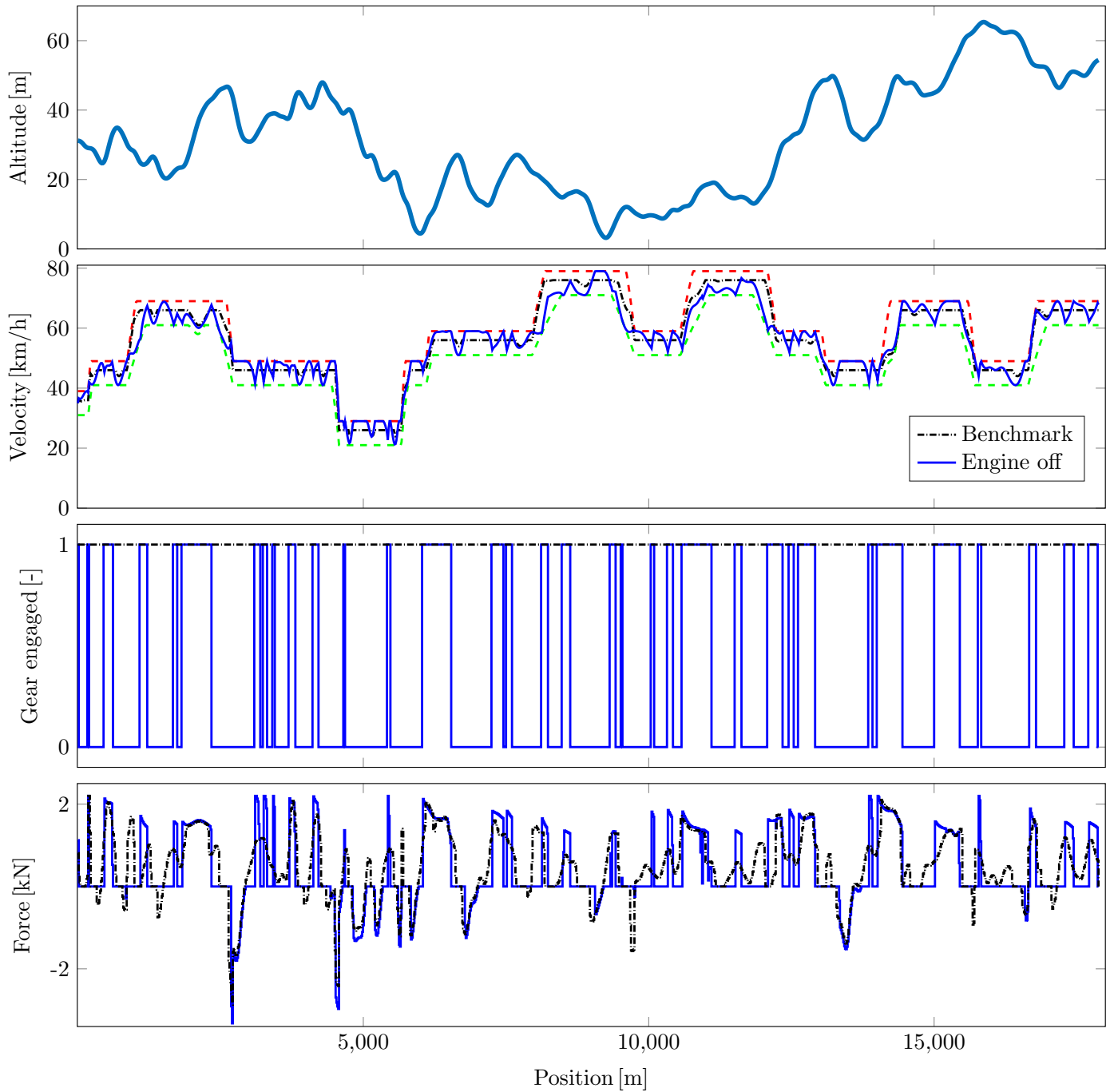}}
	\caption{Simulation results showing road altitude, velocity, gear engaged, and control force.}
	\label{fig:freewheeling_trajectories}
\end{figure*}

The savings when allowing freewheeling comes from reduction of engine drag. As can be seen, the losses due to braking actually increases when freewheeling with engine off compared to freewheeling with idling. This is because when freewheeling with idling, it is beneficial to stop freewheeling if braking is necessary. With the engine off on the other hand, the vehicle may continue to freewheel when braking, in order to avoid the penalty for gear changes. In the end, the sum of losses from braking and from engine drag are approximately the same for the two idling policies. A summary of the resulting energy consumption together with the corresponding trip time can be seen in Table~\ref{table:result}.\par

The resulting trajectories can be seen in Fig.~\ref{fig:freewheeling_trajectories} for the benchmark and for freewheeling with engine off. It can be seen that by using the latter control policy, the vehicle lowers the velocity ahead of downhills in order to avoid braking. The difference with the benchmark can be seen in the force plot at three locations during the first two kilometres. Energy is also saved by using PnG which can be seen in the frequent switching in the gear plot, even at locations without significant changes in altitude.\par   

\section{Conclusions}\label{sec:conclusion}
The driving mission of a distribution vehicle is formulated as an OCP on MIQP form. The fuel savings for the four different control policies are: 10.3\,\% for look-ahead control without freewheeling, 15.7\,\% for freewheeling with the engine idling, and 19.6\,\% for freewheeling with the engine turned off. These results indicate great potential in fuel savings in distribution vehicles by using look-ahead control. However, the fuel savings in real driving might be less due to simplifications in the vehicle model and the presence of other traffic participants.


\bibliography{library, manual}
\end{document}